\documentclass[fleqn,10pt]{wlscirep}
\usepackage[utf8]{inputenc}
\usepackage[T1]{fontenc}
\usepackage{lineno}
\usepackage{amsmath}
\graphicspath{./}
\newcommand{\beq}{\begin{equation}}
\newcommand{\eeq}{\end{equation}}
\newcommand{\bea}{\begin{eqnarray}}
\newcommand{\eea}{\end{eqnarray}}

\def\br{{\mathbf{r}}}

\title{Potential-Barrier Affinity Effect in Solid Systems}

\author[1,*]{Qiang Xu}
\author[2,*]{Zhao Liu}
\author[3,1,*]{Yanming Ma}
\affil[1]{Key Laboratory of Material Simulation Methods \& Software of Ministry of Education, College of Physics, Jilin University, Changchun, 130012, China}
\affil[2]{Institute of High Pressure Physics, School of Physical Science and Technology, Ningbo University, Ningbo, 315211, China}
\affil[3]{Center for High-Pressure Science and Technology, School of Physics and Institute of Fundamental and Transdisciplinary Research, Zhejiang University, Hangzhou, 310027, China.}

\affil[*]{e-mail: xuq@jlu.edu.cn; liuzhao@nbu.edu.cn; mym@calypso.cn}
\begin{abstract}
Electron accumulation in interatomic regions is a fundamental quantum phenomenon dictating chemical bonding and material properties, yet its origin remains elusive across disciplines\cite{sterling2024chemical,rousseau2008interstitial,miao2014high,mao2015high}. Here, we report a quantum accumulation effect---potential-barrier affinity (PBA)---revealed by solving the Schr\"odinger equation for a crystalline potential. PBA effect drives significant interatomic electron accumulation when electron energy exceeds the barrier maximum. This effect essentially enhances interatomic electron density, governing microstructures and properties of condensed matter. Our theory overturns the traditional wisdom that the interstitial electron localization in electride requires potential-well constraints\cite{rousseau2008interstitial,miao2014high,mao2015high} or hybrid orbitals\cite{racioppi2023electride}, and it serves as the fundamental mechanism underlying the formation of conventional solid bonding. This work delivers a paradigm shift in understanding electron distribution and establishes a theoretical foundation for the microscopic design of material properties.
\end{abstract}
\begin{document}
%\linenumbers

\flushbottom
\maketitle
\thispagestyle{empty}

\section*{Introduction}
As one of the most fundamental quantum phenomena, the complex behavior of electrons in interatomic region enables diverse bonding types, which govern the structures and properties of materials, making it a cornerstone of condensed matter physics, theoretical chemistry, and materials science. In particular, the interatomic electron accumulation with diverse density patterns---such as the conventional covalent bonds\cite{langmuir1919arrangement} and the unique interstitial anion electrons (IAEs) found in electrides\cite{landers1981temperature,lee2013dicalcium,dong2017electrides,miao2014high,mao2015high}---provides a versatile platform for regulating and designing material properties. For instances, diamond, the hardest known natural material, is renowned for its strong covalent bonding feature\cite{irifune2003ultrahard}; the first room-temperature air-stable inorganic electride, Ca$_{24}$Al$_{28}$O$_{64}$\cite{matsuishi2003high}, was experimentally confirmed to exhibit the IAEs, with a measured superconducting transition temperature of 0.4 K\cite{miyakawa2007superconductivity}; and electrides Cs$_3$O and Ba$_3$N possess one-dimensional IAEs together with strong interrod interactions induce band inversion in a two-dimensional superatomic triangular lattice, resulting in Dirac node lines\cite{park2018first}. Although these interatomic electron accumulation patterns govern remarkable functional properties in materials, the fundamental origin of chemical bonding remains a subject of active debate\cite{rioux1997kinetic,nordholm2020basics,levine2020clarifying,martin2022role,bacskay2022orbital}. Emerging research suggests it constitutes a complex electronic behavior governed by a confluence of mechanisms, including node-induced electron confinement, Pauli repulsion, orbital contraction, and polarization\cite{sterling2024chemical}. In electrides, conventional understanding holds two divergent views on the source of IAEs: the first explains them as confined bound states within interatomic potential wells\cite{rousseau2008interstitial,miao2014high,mao2015high,dong2017electrides}, and the second as a consequence of multicentered bonding induced by the overlap of hybrid orbitals\cite{racioppi2023electride}.

\section*{Results}
%\subsection*{The abnormal near-free electrons in electrides}
\subsection*{Abnormal interstitial electron accumulation} 
\begin{figure}[h]
\centering
\includegraphics[width=0.8\textwidth]{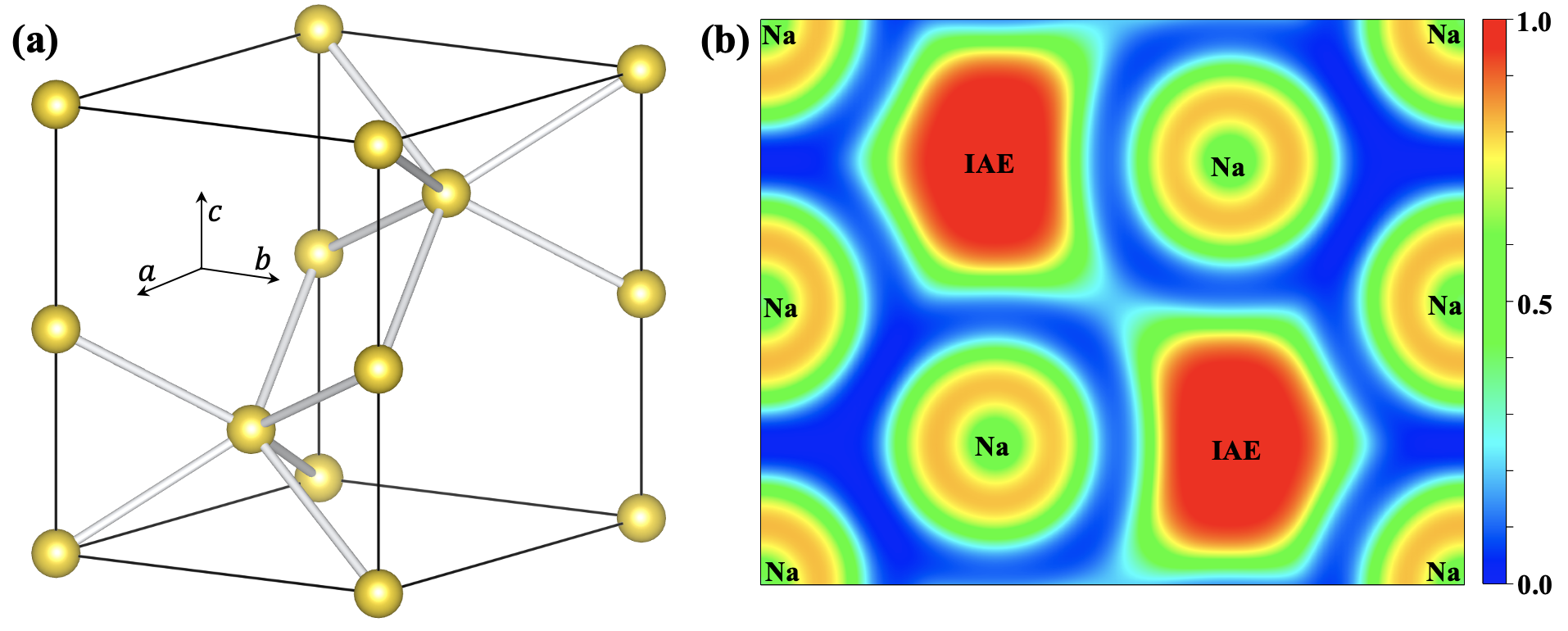}
\caption{\textbf{Structure and electronic localization function of Na-hP4.} \textbf{a,} Structure of Na-hP4 (space group P6$_3$/mmc) at 320 GPa. Lattice parameters are $a=b=2.784$~\AA~and $c=3.873$~\AA~with two inequivalent atomic positions of $2a(0,0,0)$ and $2d(2/3,1/3,1/4)$. \textbf{b,} Electron localization function of Na-hP4 plotted in the (110) plane at 320 GPa.}\label{fig1}
\end{figure}

\begin{figure}[h]
\centering
\includegraphics[width=0.8\textwidth]{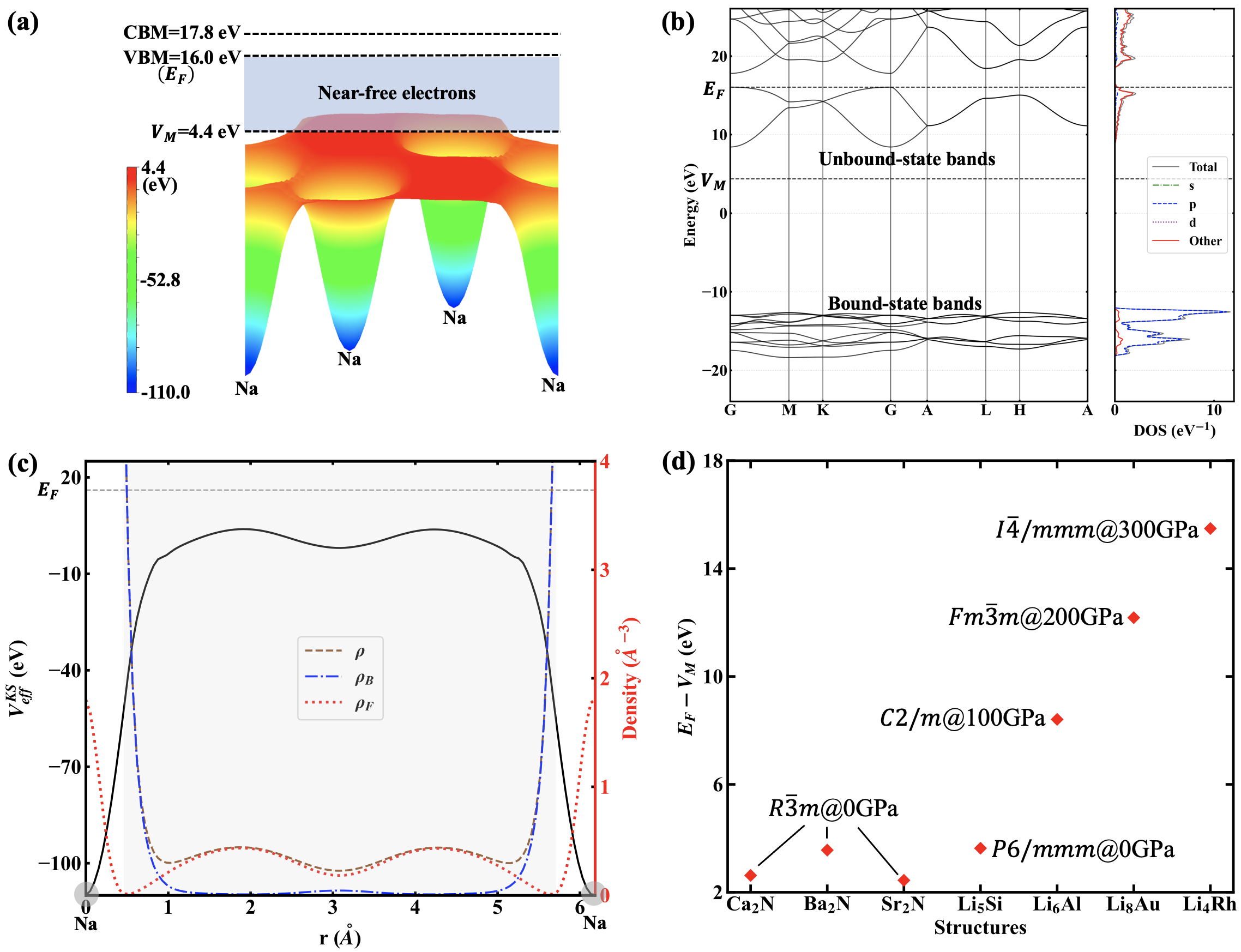}
\caption{\textbf{Evidence of the existence of near-free-valence electrons.} \textbf{a,} The Kohn-Sham effective local potential in the (110) plane of Na-hP4. \textbf{b,} Band structure (left) and density of states (right) of Na-hP4. \textbf{c,} The effective local potential $V_{eff}^{KS}$, total electron density($\rho$), bound-state derived electron density ($\rho_B$), and near-free-electron density ($\rho_F$) along [1$\bar{1}$1] direction of Na-hP4, where ``grey coverage'' denotes the interatomic region. \textbf{d,} Energy differences between Fermi energy level and maximum point of Kohn-Sham effective local potential ($E_F-V_M$) for seven typical electrides.}\label{fig2}
\end{figure}

The transparent dense sodium has been experimentally confirmed and is recognized as a typical electride\cite{ma2009transparent,mao2015high}, which has attracted much attention due to the anomalous metal-insulator phase transition under high pressure. In this study, we conducted first-principles calculations on the electronic structure of Na-hP4 at 320 GPa\cite{ma2009transparent}, as shown in Figure \ref{fig1}a, to reveal its anomalous characteristics, including the presence of a finite bandgap ($\sim$1.75 eV) and the formation mechanisms of IAEs. The electron localization function depicted in Figure \ref{fig1}b clearly shows the highly localized features of the electrons in the interstitial regions, which are consistent with the results of the previously reported differential charge density\cite{ma2009transparent}. The computational details can be found in Methods section.

We subsequently calculate the effective local potential by the Kohn-Sham density functional theory\cite{kohn1965self} in Figure \ref{fig2}a, along with the valence band maximum (VBM) and conduction band minimum (CBM). Contrary to the conventional expectation of potential wells, we found that electrons localize at the maximum point of the potential ($V_M$) within the interstitial region. Furthermore, we observed that the Fermi energy level ($E_F=16.0$ eV) is much higher than the maximum point of the potential ($V_M=4.4$ eV), indicating that near-free electrons (NFEs) occupy the single-particle states with energy levels ranging from $V_M$ to $E_F$. The band structure shown in Figure~\ref{fig2}b (left) displays two valence electron bands (4 electrons per cell) located within the NFE energy range. This suggests that all valence electrons behave as unbound-state NFEs rather than being confined as interstitial bound electrons. While early models interpreted the interstitial-potential well as arising from Pauli exclusion by core electrons\cite{neaton2001constitution,rousseau2008interstitial}, a quantitatively accurate description of the Pauli potential remains absent within the fermion quasiparticle framework. We address this point further in the Method section.

Additionally, the density of states (DOS) in Figure~\ref{fig2}b (right) demonstrates that the \textit{s}, \textit{p}, and \textit{d} components are negligible compared to the total DOS within the valence-electron energy range, which is also a distinct feature commonly associated with unbound-state NFEs\cite{liu2025mechanism}, rather than the multicentered bonding of hybrid orbitals. In addition, the concept of hybrid orbital derives from a linear combination of localized atomic orbitals centered on a single atom site\cite{pauling1931nature}. As such, it should be applied with caution when describing unbound states. To further clearly demonstrate the presence of the electron localization within the interstitial region, we plot the Kohn-Sham effective local potential ($V_{eff}^{KS}$), total ($\rho$), bound-state ($\rho_B$), and near-free ($\rho_F$) electron densities along the [1$\bar{1}$1] direction of Na-hP4 in Figure~\ref{fig2}c. We observed that the NFE bands contribute almost all the interstitial electron density near the maximum potential point, while the electron density tails from the bound-state bands can be ignored. This result directly confirms that interstitial electrons originate from near-free-valence states. We also present the effective ionic potential of Na$^+$-hP4 in Figure~5 of the Method Section by removing the valence electrons. The analysis further confirms that IAEs behave as NFEs with high density accumulation around the potential maximum. Notably, we further found evidence for the existence of NFEs ($E_F>V_M$) in representative electrides\cite{qiu2022superconductivity,wang2023crystal,lee2013dicalcium,you2022emergent,wang2022pressure,zhang2023superconductivity,guan2025predicted}, as illustrated in Figure \ref{fig2}d, suggesting that the existence of near-free valence electrons is a common phenomenon among electrides. More evidence of the NFE-density distribution in the interstitial region can be found in Figure~6 and Figure~7 of the Method section. These findings compel a reevaluation of the current understanding of IAEs in electrides.

\begin{figure}[h]
\centering
\includegraphics[width=0.8\textwidth]{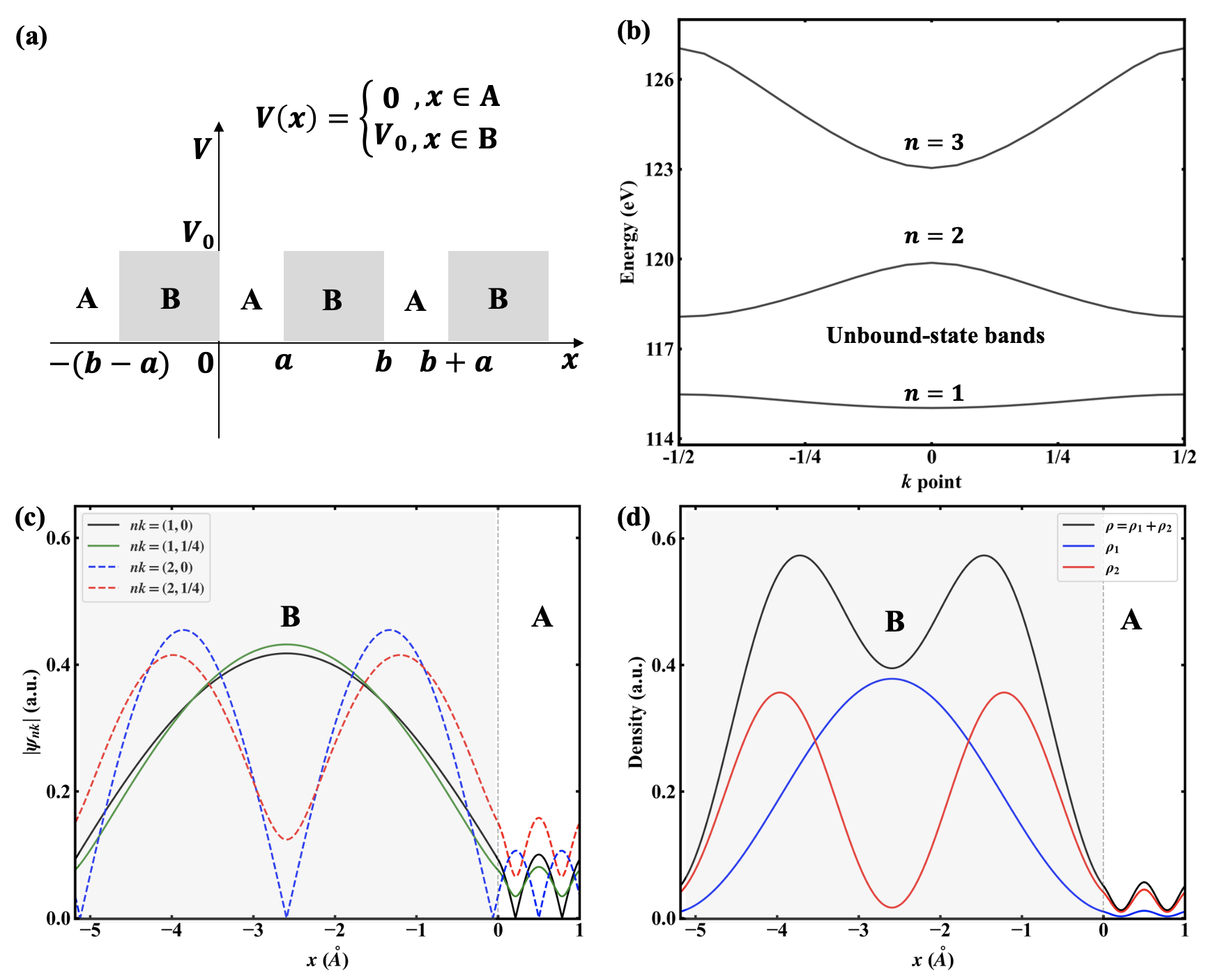}
\caption{\textbf{Unbound-state Kronig-Penney model.} \textbf{a,} Periodic square potential well and barrier model, where ``A'' and ``B'' denote the atomic and interatomic regions; ``$V_0$'' is the height of the square potential barrier; ``$a$'' and ``$b$'' are the potential-well and primary-cell width, respectively. \textbf{b,} near-free-electron band structure of Kronig-Penney model. \textbf{c,} Absolute wavefunctions of $n=1$ and $n=2$ bands. \textbf{d,} Electron density calculated for $n=1$ band ($\rho_1$) and $n=2$ band ($\rho_2$), as well as the total density $\rho=\rho_1+\rho_2$, where ``grey coverage'' denotes the ``B'' region. The used parameters of the Kronig-Penney model are $a=1.0$\AA, $b=6.185$\AA, and $V_0=114.0$ eV.}\label{fig3}
\end{figure}

\subsection*{Potential-barrier affinity effect}

To explore the nature of NFE states ($\epsilon_{nk}>V_0$) in a periodic field, as shown in Figure \ref{fig3}a, we establish a one-dimensional generalized Kronig-Penney (KP) model for unbound states, providing a direct counterpart to the traditional bound-state scheme\cite{kronig1931quantum}, where $\epsilon_{nk}$ indicates the eigenvalue of $nk$th Bloch state ($\psi_{nk}$); ``A'' and ``B'' denote the atomic and interatomic regions, respectively. The detailed information of KP model in this work can be found in Method Section. According to the Schr\"odinger equation, the wavefunctions of unbound states have a general form:
\bea
\psi_{nk}^\text{A}(x)=A_{nk}e^{i\alpha_{nk} x}+A'_{nk}e^{-i\alpha_{nk} x},~x\in \text{A};\nonumber\\
\psi_{nk}^\text{B}(x)=B_{nk}e^{i\beta_{nk} x}+B'_{nk}e^{-i\beta_{nk} x},~x\in \text{B},\label{eq:1}
\eea
\noindent where $\psi^\text{A}_{nk}$ and $\psi^\text{B}_{nk}$ are the branches of the wavefunction in regions A and B, respectively; $A_{nk}$, $A_{nk}'$, $B_{nk}$, and $B_{nk}'$ are the corresponding plane-wave-component coefficients in the wavefunction. The wavevectors ($\alpha_{nk}$ and $\beta_{nk}$) in A and B regions are $\alpha_{nk}=\sqrt{2\epsilon_{nk}}$ and $\beta_{nk}=\sqrt{2(\epsilon_{nk}-V_0)}$, respectively. The details of the numerical solution for the Schr\"odinger equation of KP model can be found in the Method Section. In Figure \ref{fig3}b, we present the unbound-state band structure ($\epsilon_{nk}>V_0$) under the KP potential, which shows finite band gaps and parabolic characteristics as the perturbed free-electron gas\cite{simon2013oxford,girvin2019modern}. The concavities of the unbound-state bands in Figure \ref{fig3}b are also similar to those in Figure \ref{fig2}b for Na-hP4 near the $G$ point.

Subsequently, we calculated the absolute value of the wavefunctions and the electron density in Figure \ref{fig3}c and Figure \ref{fig3}d, respectively, corresponding to the $n=1,2$ bands in Figure \ref{fig3}b. All of them exhibit counterintuitive phenomena, where the large amplitudes are observed in the interatomic region. This is consistent with the electron density of Na-hP4 in Figure \ref{fig2}c. Furthermore, we reproduced the double peaks of the interatomic electron density in Figure \ref{fig2}c by our unbound-state KP model. Figure \ref{fig3}c shows that the band index ($n$) is directly related to the number of peaks of $|\psi_{nk}(x)|$ in the barrier region. Herein, we do not consider the electron density in the atomic region because of the large differences in the atomic central potential and the square potential. Instead, we will mainly use the unbound-state KP model to demonstrate that the steep-shaped potential can lead to a high electron density in the interatomic region. In the unbound-state regime ($\epsilon_{nk}>V_0$), we can directly find that the wavefunction exhibits the properties of short and long wavelengths in regions A (atomic) and B (interatomic) regions, respectively, because of the inequality $\alpha_{nk}>\beta_{nk}$. Therefore, the wavefunction-continuity conditions at the point of barrier-well junction, such as $x=0$, can be written as:
\bea
\psi_{nk}^\text{A}(0)=\psi_{nk}^\text{B}(0);\nonumber\\
\left.\frac{d\psi_{nk}^\text{A}}{dx}\right|_{x=0}=\left.\frac{d\psi_{nk}^\text{B}}{dx}\right|_{x=0}.\label{eq:2}
\eea
The continuity conditions in Eq.~(\ref{eq:2}) ensure the wavefunction smoothness at $x=0$. In particular, the first-derivative continuity of the wavefunction leads to the amplitude of the long-wavelength plane-wave component being greater than that of the short-wavelength one, resulting in the PBA effect with high electron density distribution in the interatomic region. Quantitatively, the following equations can be derived for the coefficients by Eqs.~(\ref{eq:1}--\ref{eq:2}):
\bea
B_{nk}=\frac{A_{nk}+A_{nk}'}{2}+\frac{1}{\sqrt{1-V_0/\epsilon_{nk}}}\frac{A_{nk}-A_{nk}'}{2}=\bar{A}_{nk}+\lambda_{nk}\Delta_{nk};\nonumber\\
B_{nk}'=\frac{A_{nk}+A_{nk}'}{2}-\frac{1}{\sqrt{1-V_0/\epsilon_{nk}}}\frac{A_{nk}-A_{nk}'}{2}=\bar{A}_{nk}-\lambda_{nk}\Delta_{nk}.\label{eq:3}
\eea
With Eq.~(\ref{eq:3}), we found that the coefficient $B$ and $B'$ are determined by the average $\bar{A}_{nk}=(A_{nk}+A_{nk}')/2$ and difference $\Delta_{nk}=(A_{nk}-A_{nk}')/2$ of coefficients $A_{nk}$ and $A_{nk}'$, as well as the factor $\lambda_{nk}=(\sqrt{1-V_0/\epsilon_{nk}})^{-1}$. For the special case, $\Delta_{nk}=0$ for $A_{nk}=A_{nk}'$, the plane-wave components share the same amplitudes $|A_{nk}|=|A_{nk}'|=|B_{nk}|=|B_{nk}'|$. For the general cases, $\Delta_{nk}\neq 0$, $V_0/\epsilon_{nk}$ is an important quantity that determines the amplitudes of the plane-wave components. When the energy level is much higher than the potential barrier ($\epsilon_{nk}>>V_0$, or $\lambda_{nk}\to 1$), the difference in the component amplitudes between A and B regions can be ignored; Conversely, if the energy level approaches the height of the potential barrier ($\epsilon_{nk}\to V_0$ and $\lambda\to\infty$), the unbound-sate wavefunction exhibits a high component characteristic in the barrier region:
\bea
\lim_{\epsilon_{nk} \to V_0}|B_{nk}|,|B_{nk}'|\sim  \lambda_{nk}|\Delta_{nk}|>>|A_{nk}|,|A_{nk}'|.\label{eq:4}
\eea
So far, we demonstrate that the wavefunctions derived from Eq.~(\ref{eq:4}), with high-component amplitudes within interatomic region, will result in a high interatomic electron density (PBA effect). Therefore, we immediately draw the conclusion: if there are sufficiently large occupied states (or DOS) near the potential barrier ($V_0$), the system can exhibit excellent PBA electrons, such as the IAEs in electrides. Here, $V_0$ can be regarded as $V_M$ in Figure~\ref{fig2}a when the lowest potential point is shifted to 0 eV.

According to the results of the KP model, we find that the PBA of quantum state is dramatically consistent with the classical physical intuition: a free particle has an initial kinetic energy $T_0$; When it passes through the barrier region $V(x)>0$, part of the kinetic energy is converted into potential energy, and its velocity decreases due to the smaller kinetic energy $T(x)=T_0-V(x)<T_0$, causing it to have a higher probability of being detected in the higher barrier region. When the initial kinetic energy is much higher than the potential barrier $T_0>>V(x)$, the particle can move freely throughout space $T(x)\approx T_0$. This finding seems to indicate that high-energy PBA electrons tend to classical behavior according to Bohr's correspondence principle\cite{tipler2007modern}.

\begin{figure}[h]
\centering
\includegraphics[width=0.8\textwidth]{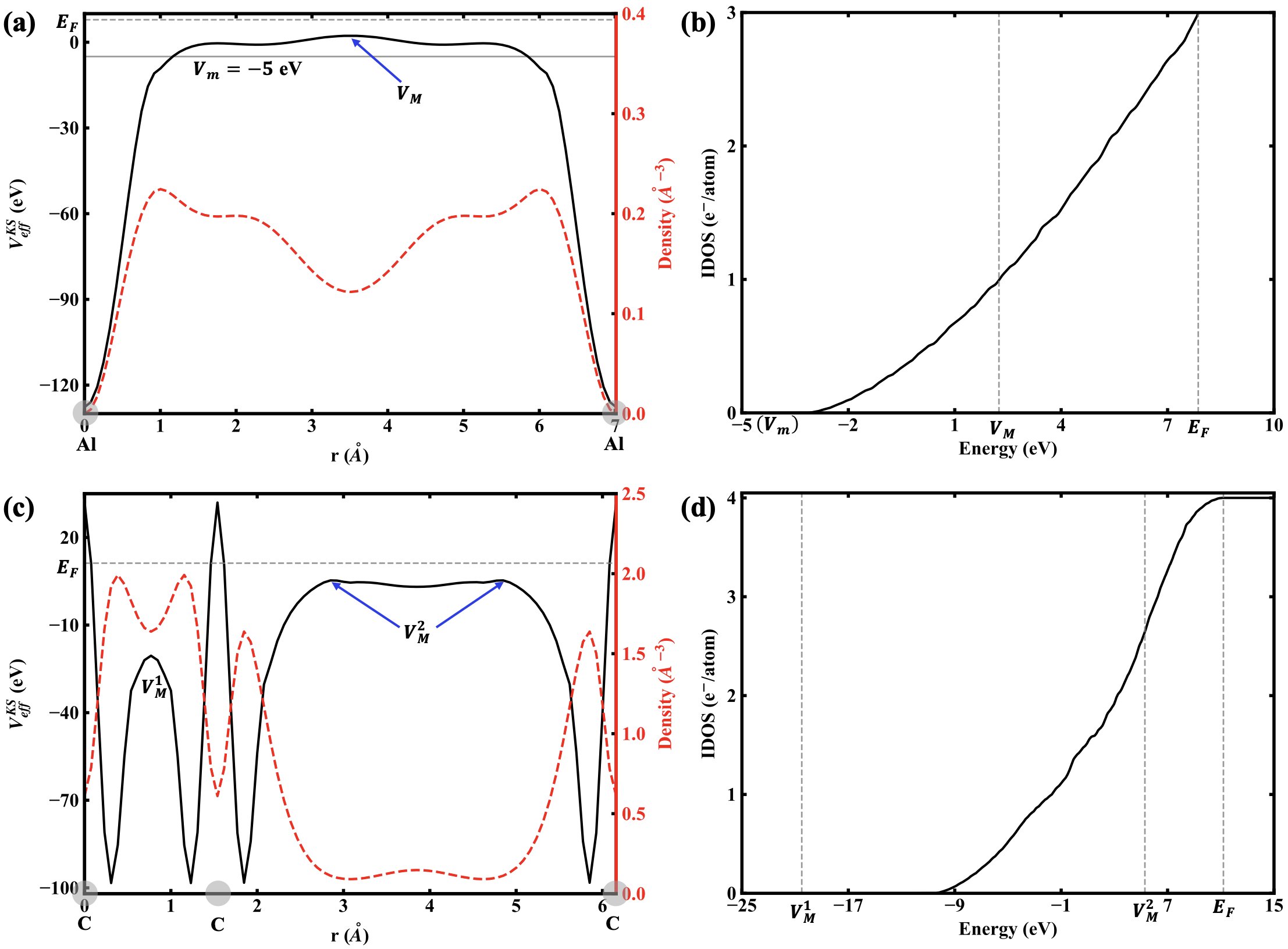}
\caption{\textbf{Potentials and densities of the systems containing metallic and covalent bonds.} The Kohn-Sham effective local potentials and electron densities along [111] direction of \textbf{a,} Face-centered-cubic (FCC) Al, and \textbf{c,} Diamond structures. The calculated integrated density of states, IDOS($E$)=$\int_{-\infty}^E$DOS($\epsilon$)$d\epsilon$ of \textbf{b}, FCC-Al and \textbf{d,} Diamond. $E_F$, $V_m$, $V_M$, and $V_M^n$ denote the Fermi energy level, the potential barrier edge, the global and $n$th-local maximum potential points, respectively.
}\label{fig4}
\end{figure}

\subsection*{Metallic and covalent bonds}
The PBA effect offers a natural starting point for reexamining the formation of metallic bonding, given the inherent abundance of NFEs in metals. Therefore, we calculated the Kohn-Sham effective local potentials and electron densities for aluminum (Al). As shown in Figure~\ref {fig4}a and Figure~\ref {fig4}(b), Al metal contains numerous near-free-valence electrons ($V_M<\epsilon_{nk}<E_F$). Crucially, all these valence electron energy levels lie above the potential barrier edge ($\epsilon_{nk}>V_m=-5$~eV), forming a high electron density in the interatomic region. While a few energy levels, as shown in Figure~\ref{fig4}b, are lower than the barrier maximum ($V_M$), resulting in their electron density tails decaying in the interatomic region. Furthermore, the relative alignment between the local potential barrier height and electronic energy levels suggests that the PBA effect may also contribute to the formation of certain covalent bonds. As the results of diamond structure shown in Figure~\ref {fig4}c, neighboring atoms are covalently bonded, while all the valence-band energies are higher than the local barrier ($V_M^1<\epsilon_{nk}<E_F$) in Figure~\ref {fig4}d in the interatomic bonding region($\br\approx 0.8$\AA), resulting in a high interatomic electron density as covalent bond. For the local barrier region on the right side (2.3\AA$<\br<5.5$\AA), there are a few bands higher than the barrier ($V_M^2<\epsilon_{nk}<E_F$) in Figure~\ref {fig4}d, resulting in a small peak at $\br\approx3.8$\AA. To clarify the PBA contribution of the local potential barrier in solid bonding, we also propose a more generalizable extension of the KP model, which can be implemented as a double-barrier configuration, as illustrated in Figure~8. Importantly, we employ the generalized KP model to provide a clear definition of solid bonds within the quasiparticle picture. The detailed discussion can be found in the Method section. Consequently, our results establish the PBA phenomenon as a universal rule governing the behavior of higher-energy electrons traversing (local) barriers in the interatomic regions, adding a unique perspective for the formation of both metallic and covalent bonds in condensed matter.

\section*{Discussion}
In summary, we theoretically reveal a universal PBA effect by solving the Schr\"odinger equation of a one-dimensional crystal potential. The PBA effect is characterized by significant electron accumulation within the interatomic region, arising from the smooth connection of wavefunctions between short-wavelength components in the atomic region and long-wavelength components across the interatomic region, amplifying the latter and producing electron accumulation there. We demonstrate that the PBA effect becomes most pronounced as the electron energy approaches the barrier maximum and fundamentally governs the resulting electron density distribution in condensed matter, thereby dictating their microstructures and properties. From the PBA effect, we derive the necessary conditions for the existence of electrides: (i) the Fermi level lies above the maximum of the effective potential barrier; (ii) the barrier maximum is located in a spatially open region far from nuclei; and (iii) an appreciable density of occupied states exists between the Fermi level and the barrier maximum. Such conditions provide a physically transparent guideline for the discovery of an electride. Furthermore, this theory overturns the traditional belief that the formation of IAEs strictly requires potential well constraints or multicentered bonding, and it serves as the fundamental mechanism underlying the formation of conventional solid bonding, defect states, and excited-state electron density distribution. This unexpected discovery represents a fundamental paradigm shift in understanding electron behavior in materials, providing a universal theoretical foundation for the microscopic design of specific material properties.

\section*{Methods}
\subsection*{First-principles calculations}
The first-principles electronic structures, including effective local potential, electron localization function, electron density, and band structure, are carried out using the Kohn-Sham density functional theory (KSDFT)\cite{kohn1965self} as implemented in the Vienna Ab initio Simulation Package (VASP v6.4.1)\cite{kresse1996efficient}. The Perdew-Burke-Ernzerhof (PBE) exchange-correlation functional\cite{perdew1996generalized} and the projected-augmented-wave method\cite{blochl1994projector} are adopted for all KSDFT calculations. The kinetic energy cutoff of 1000 eV for the plane-wave basis set and $k$-spacing of 0.15~\AA$^{-1}$ within Monkhorst-Pack sampling\cite{monkhorst1976special} are used for all cases. The dense real-space grids of $120\times 120\times 160$ are adopted for calculating the high-resolution electron localization function in Figure \ref{fig1}b. Notably, the crystal structures of Na\cite{ma2009transparent}, Ba$_2$N\cite{qiu2022superconductivity}, Ca$_2$N\cite{lee2013dicalcium}, Li$_5$Si\cite{you2022emergent}, Li$_6$Al\cite{wang2022pressure}, Li$_8$Au\cite{zhang2023superconductivity} , Li$_4$Rh\cite{guan2025predicted} utilized in this study were obtained from previously published sources; the structure of diamond and Al come from the built-in structure library of Material Studio.
\begin{figure}[htbp]
\centering
\includegraphics[width=0.8\textwidth]{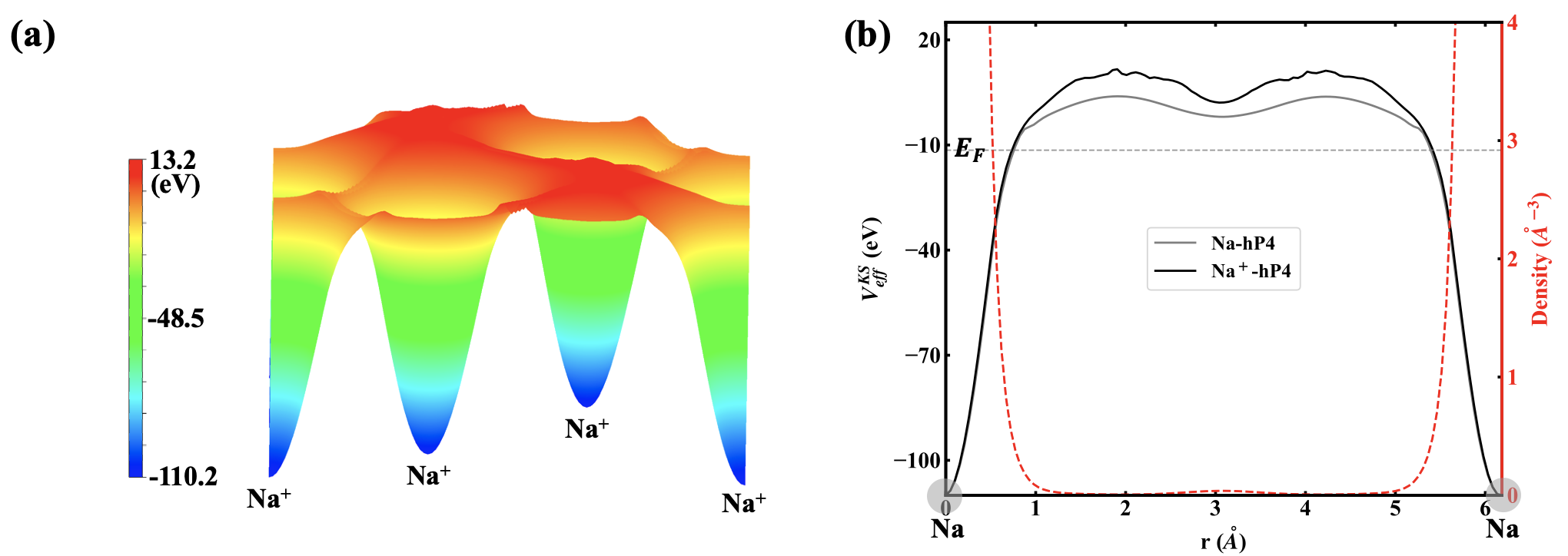}
\caption{The Kohn-Sham effective potential and electron density of Na$^+$-hP4 in (a) (110) plane, and (b) [1$\bar{1}$1] direction. The grey line denotes the reference potential of Na-hP4.}\label{figs1}
\end{figure}

To isolate the interatomic potential field from the screening effect of valence electrons, we calculated the KSDFT-derived effective potential for a hypothetical system of Na$^+$-hP4 by removing the valence electrons out of Na-hP4. As shown in Figure~\ref{figs1}(a) and (b), the maximum of this ``ion-only'' potential remains located at the interstitial site. This is expected, as the potential converges toward the superposition of screened ionic Coulomb potentials in the interatomic region. Notably, the absence of near-free electron screening leads to a higher interatomic potential barrier in Na$^+$-hP4 than in Na-hP4. Crucially, this independent calculation confirms our central thesis: interstitial anionic electrons (IAEs) are near-free electrons with high density accumulation around the potential maximum, not bound states confined to potential wells.

\subsection*{Near-free electrons in electrides}

\begin{figure}[htbp]
\centering
\includegraphics[width=0.7\textwidth]{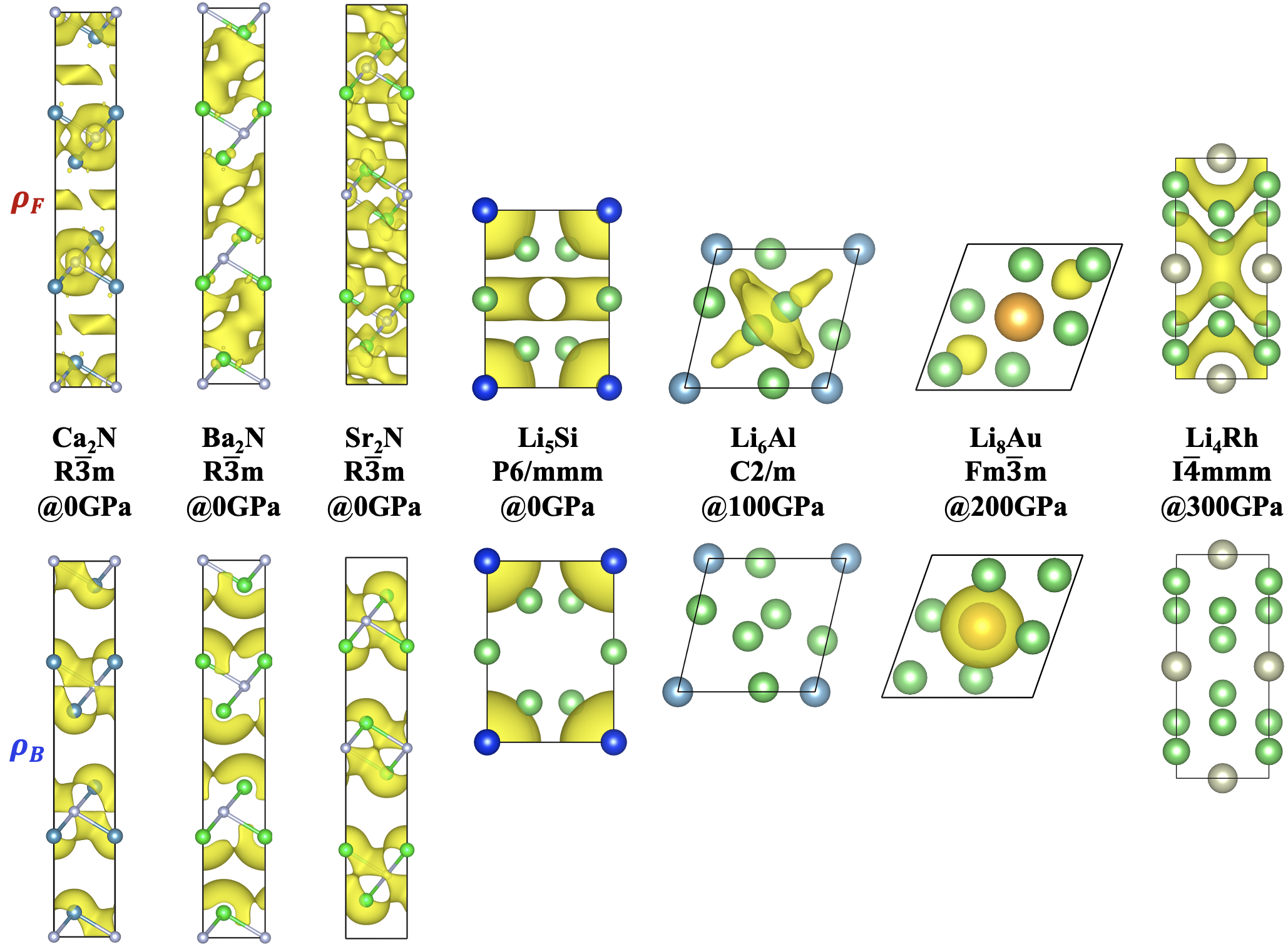}
\caption{The electron density distribution functions for near-free electrons ($\rho_F$) and bound-state electrons ($\rho_B$), where $\rho_F$ and $\rho_B$ are calculated according to the unbound-state bands ($V_M<\epsilon_{nk}<E_F$) and bound-state bands ($\epsilon_{nk}<V_M$), respectively. The isosurfaces of 0.008, 0.005, 0.008, 0.01, 0.04, 0.1, and 0.05 \AA$^{-3}$ are used for Ca$_2$N, Ba$_2$N, Sr$_2$N, Li$_5$Si, Li$_6$Al, Li$_8$Au, and Li$_4$Rh, respectively.}\label{figse1}
\end{figure}

\begin{figure}[htbp]
\centering
\includegraphics[width=0.7\textwidth]{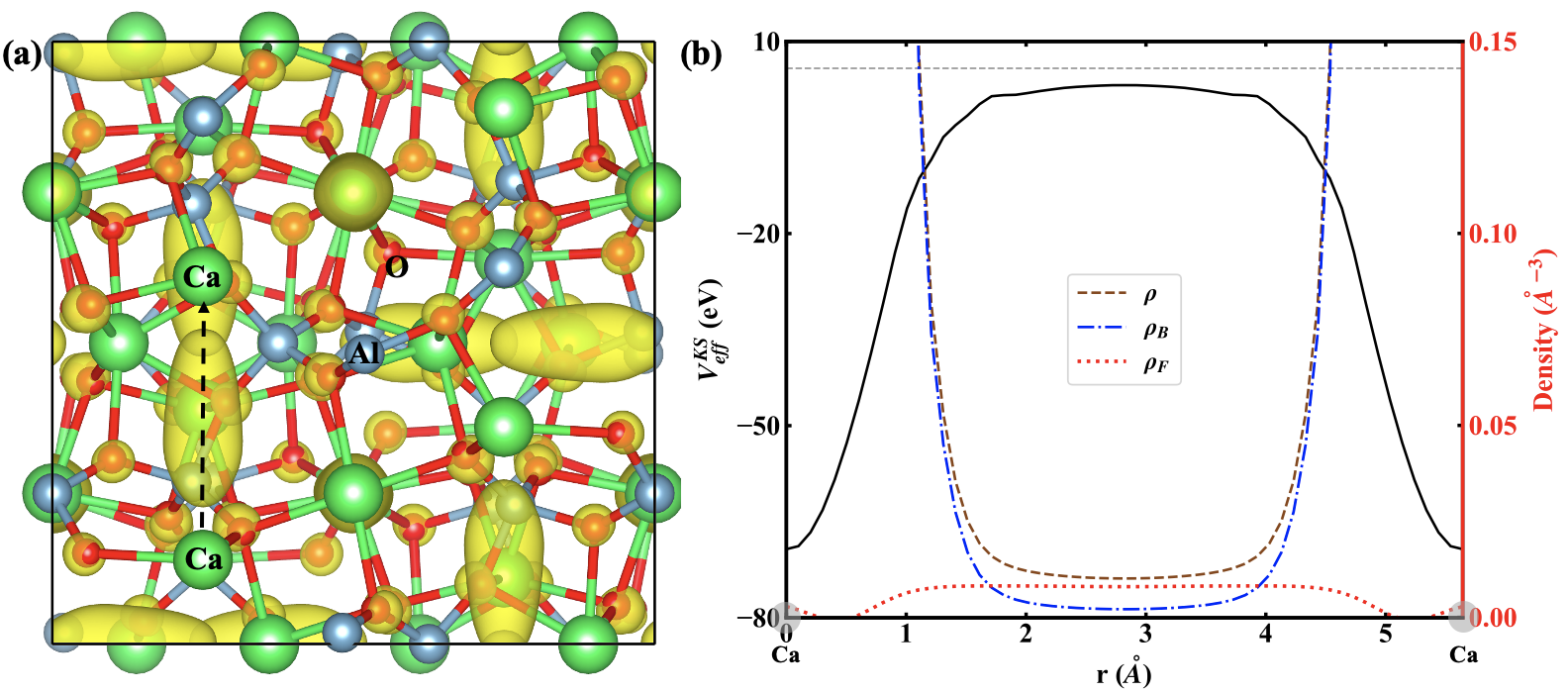}
\caption{(a) The electron density distribution functions for near-free electrons ($\rho_F$), and (b) the effective potential and electron density calculated along the dashed line of the Ca-Ca pair for Ca$_{24}$Al$_{28}$O$_{64}$, where the isosurface is 0.001 \AA$^{-3}$.}\label{figse2}
\end{figure}

To further establish that IAEs arise from near-free-electron (NFE) behavior, we calculated the partial electron density from band decomposition for the electrides in Fig.~2(d) of the maintext, as well as the representative mayenite [Ca$_{24}$Al$_{28}$O$_{64}$]\cite{matsuishi2003high}. The results, presented in the Figure~\ref{figse1} and Figure~\ref{figse2}, reveal that the NFE density accumulates predominantly in the interstitial region, in agreement with the findings for Na-hP4.

\subsection*{Unbound-state solver for Kronig-Penney potential}
Solving the unbound states within the KP potential is the foundation for understanding the anomalous PBA effect. Here, we provide the complete details to solve the Schr\"odinger equation:
\bea
\frac{1}{2}\frac{d^2\psi_{nk}(x)}{dx^2}+[\epsilon_{nk}-V(x)]\psi_{nk}(x)=0,\label{eq:5}
\eea
\noindent where $V(x)$ is the periodic KP potential\cite{kronig1931quantum} as shown in Figure \ref{fig3}a. The unbound-state wavefunction can be written in a direct sum of regions A and B:
\bea
\psi_{nk}(x)=\psi_{nk}^\text{A}(x)\oplus\psi_{nk}^\text{B}(x),\label{eq:6}
\eea
\noindent where the formal solutions of $\psi_{nk}^\text{A}$ and $\psi_{nk}^\text{B}$ can be found in Eq.~(\ref{eq:1}). We then can obtain the wavefunction with the periodicity of the primitive cell according to Bloch's theorem\cite{bloch1929quantenmechanik} ($u_{nk}=e^{-ikx}\psi_{nk}$):
\bea
u_{nk}^\text{A}(x)=e^{-ikx}\psi_{nk}^\text{A}(x)=A_{nk}e^{i(\alpha_{nk}-k)x}+A_{nk}'e^{-i(\alpha_{nk}+k)x},~x\in\text{A};\nonumber\\
u_{nk}^\text{B}(x)=e^{-ikx}\psi_{nk}^\text{B}(x)=B_{nk}e^{i(\beta_{nk}-k)x}+B_{nk}'e^{-i(\beta_{nk}+k)x},~x\in\text{B},\label{eq:7}
\eea
\noindent where the wavevectors $\alpha_{nk}=\sqrt{2\epsilon_{nk}}$ and $\beta_{nk}=\sqrt{2(\epsilon_{nk}-V_0)}$ represent the short and long wavelengths, respectively, as mentioned in the Results section. To obtain the coefficients ($A_{nk}$, $A_{nk}'$, $B_{nk}$, $B_{nk}'$) and eigenvalues ($\epsilon_{nk}$), we applied the continuity conditions at $x=0$ and the periodic boundary conditions at $x=a$ and $x=-(b-a)$:
\bea
u_{nk}^\text{A}(0)=u_{nk}^\text{B}(0),\nonumber\\
\left.\frac{du_{nk}^\text{A}}{dx}\right|_{x=0}=\left.\frac{du_{nk}^\text{B}}{dx}\right|_{x=0};\nonumber\\
u_{nk}^\text{A}(a)=u_{nk}^\text{B}(-(b-a)),\nonumber\\
\left.\frac{du_{nk}^\text{A}}{dx}\right|_{x=a}=\left.\frac{du_{nk}^\text{B}}{dx}\right|_{x=-(b-a)}.\label{eq:8}
\eea
By combining Eqs.~ (\ref{eq:7}), and (\ref{eq:8}), we can derive the following linear equations:
\beq
\left(
\begin{array}{cccc}
1 & 1 & -1 & -1 \\
\alpha_{nk}-k & -(\alpha_{nk}+k) & -(\beta_{nk}-k) & \beta_{nk}+k \\
e^{i(\alpha_{nk}-k)a} & e^{-i(\alpha_{nk}+k)a} & -e^{-i(\beta_{nk}-k)(b-a)} & -e^{i(\beta_{nk}+k)(b-a)} \\
(\alpha_{nk}-k)e^{i(\alpha_{nk}-k)a} & -(\alpha_{nk}+k)e^{-i(\alpha_{nk}+k)a} & -(\beta_{nk}-k)e^{-i(\beta_{nk}-k)(b-a)} & (\beta_{nk}+k)e^{i(\beta_{nk}+k)(b-a)} 
\end{array}
\right)
\left(
\begin{array}{c}
A_{nk} \\
A_{nk}' \\
B_{nk} \\
B_{nk}'
\end{array}
\right)=0.\label{eq:9}
\eeq
If Eq.~({\ref{eq:9}}) has the nonzero vector solutions, the determinant of the 4$\times$4 matrix must be zero, which leads to the following equation:
\bea
cos(kb)=-\frac{\alpha^2+\beta^2}{2\alpha\beta}sin(\alpha a)sin(\beta(b-a))+cos(\alpha a)cos(\beta(b-a)). \label{eq:10}
\eea
Given $V_0$, $a$, $b$ and $k$ point, numerically solving Eq.~(\ref{eq:10}) can obtain the eigenvalues or band structure \{$\epsilon_{nk}$\} for $\epsilon_{nk}>V_0$. Finally, we solve the standard homogeneous linear equations of Eq.~(\ref{eq:10}) based on the eigenvalues to obtain the coefficient vector ($A_{nk}$, $A_{nk}'$, $B_{nk}$, $B_{nk}'$)$^\text{T}$ for the wavefunctions in Eq.~(\ref{eq:7}) or Eq.~(\ref{eq:1}).

In this work, the parameters of the square potential barrier in KP model are adopted as close as possible to the first-principles results of Na-hP4: $V_0=114.0$~eV and $a=1.0$~\AA~are obtained from the height and half width, respectively, of $V_{eff}^{KS}$ near the atomic region; $b=6.185$~\AA~is set as the length of [1$\bar1$1] direction of Na-hP4 according to Figure \ref{fig2}c. 21 $k$ points are used for band structure and electron density calculations in the KP model.

\subsection*{Generalized doulbe-barrier Kronig-Penney model}

\begin{figure}[htbp]
\centering
\includegraphics[width=0.80\textwidth]{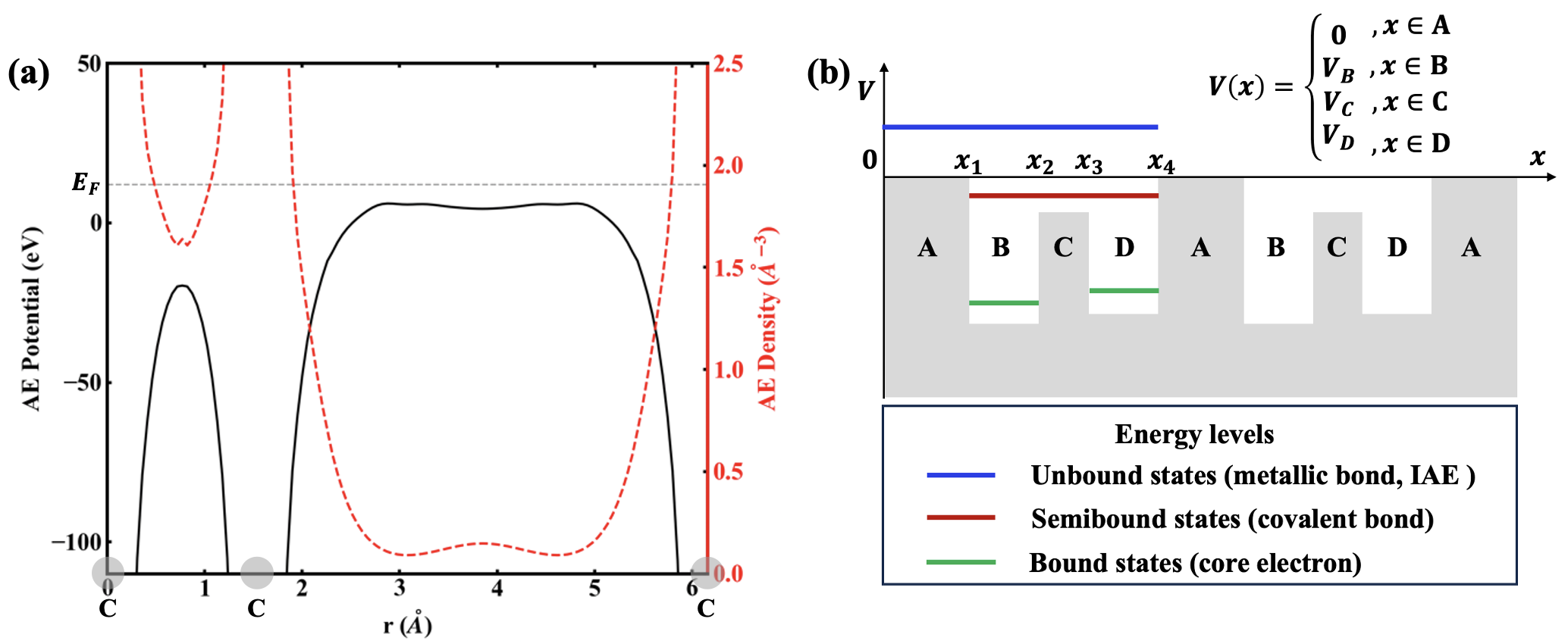}
\caption{(a) The effective potential and electron density calculated by the all-electron potential for Diamond; (b) The generalized KP model with two different barriers (or wells) in a unit cell. ``A''/``C'' and ``B''/``D'' denote the interatomic regions and atomic regions, respectively. Especially, ``C'' also indicates the covalent bonding region.
}\label{figs2}
\end{figure}

Figure 4(c) of the maintext shows that the effective potential in diamond reaches a maximum at the carbon sites. This feature originates from the pseudopotential representing the core-electron interaction, an inherent characteristic of the frozen-core approximation within the PAW method. To clarify the physical picture, we performed all-electron (AE) calculations, the results of which are presented in Figure~\ref{figs2}(a). These calculations confirm the highly localized core-electron density at the carbon nuclei and the corresponding deep effective potential wells centered on these atomic sites.

To give a physical picture of the covalent bonding [see Figure 4(c) in our maintext], we further propose a more generalizable extension of the Kronig-Penney (KP) model, which can be implemented as a double-barrier (or double-well) in a unit cell for the diamond structure, as illustrated in Figure~\ref{figs2}(b). Here, we shift the maximum potential to 0 eV for convenience.

For such a square barrier-well potential, the components of the wavefunction adopt one of two different forms depending on the electron energy: (i) an oscillatory, planewave-like solution when the energy exceeds the potential barrier, or (ii) an exponentially decaying solution when the energy is below the potential barrier. For the square barrier-well potential with the configuration of $V_B\leq V_D<V_C<V_A$ as shown in Figure~\ref{figs2}(b), the energy levels ($\epsilon_{nk}$) can be classified into three categories: (1) bound states ($V_B<\epsilon_{nk}<V_C$) for localized electrons (core states); (2) semibound states ($V_C<\epsilon_{nk}<V_A$) for covalent bonds; and (3) unbound states ($\epsilon_{nk}>V_A$) for near-free electrons, metallic bonds, or interstitial anionic electrons.

For comparison with the covalent bonds in the diamond system, our analysis focuses on the semibound state in Figure~\ref{figs2}(b) with energy in between the two barriers ($V_C<\epsilon_{nk}<V_A$ ), where the wavefunction components in different regions can be written as:

\begin{align}
\psi_{nk}^\text{A}(x)=A_{nk}e^{\alpha_{nk}x}+A'_{nk}e^{-\alpha_{nk}x},~x\in\text{A};\nonumber\\
\psi_{nk}^\text{B}(x)=B_{nk}e^{i\beta_{nk}x}+B'_{nk}e^{-i\beta_{nk}x},~x\in\text{B};\nonumber\\
\psi_{nk}^\text{C}(x)=C_{nk}e^{i\gamma_{nk}x}+C'_{nk}e^{-i\gamma_{nk}x},~x\in\text{C};\nonumber\\
\psi_{nk}^\text{D}(x)=D_{nk}e^{i\eta_{nk}x}+D'_{nk}e^{-i\eta_{nk}x},~x\in\text{D}.\label{eq:S7}
\end{align}

We emphasize that in region A, the wavefunction must decay exponentially, as indicated by the Schr\"odinger equation for a classically forbidden region. This behavior is fundamentally different from the planewave-like character of the wavefunctions in regions B, C, and D. With the eight numerical conditions (six continuity conditions at the points of $x=\{x_1, x_2,x_3\}$; two periodic boundary conditions at $x=0$ and $x=x_4$), the eight coefficients in Eq.~(\ref{eq:S7}) can be strictly solved. The potential-barrier affinity effect can occur in region C with planewave-like wavefunctions, as demonstrated in the maintext of our manuscript. We note that systematically adjusting the relative height and spatial separation of the potential barriers in this generalized model may yield deeper insights into the understanding of the formation mechanisms of different bonding types.

\subsection*{Pauli potential}

\begin{figure}[htbp]
\centering
\includegraphics[width=0.8\textwidth]{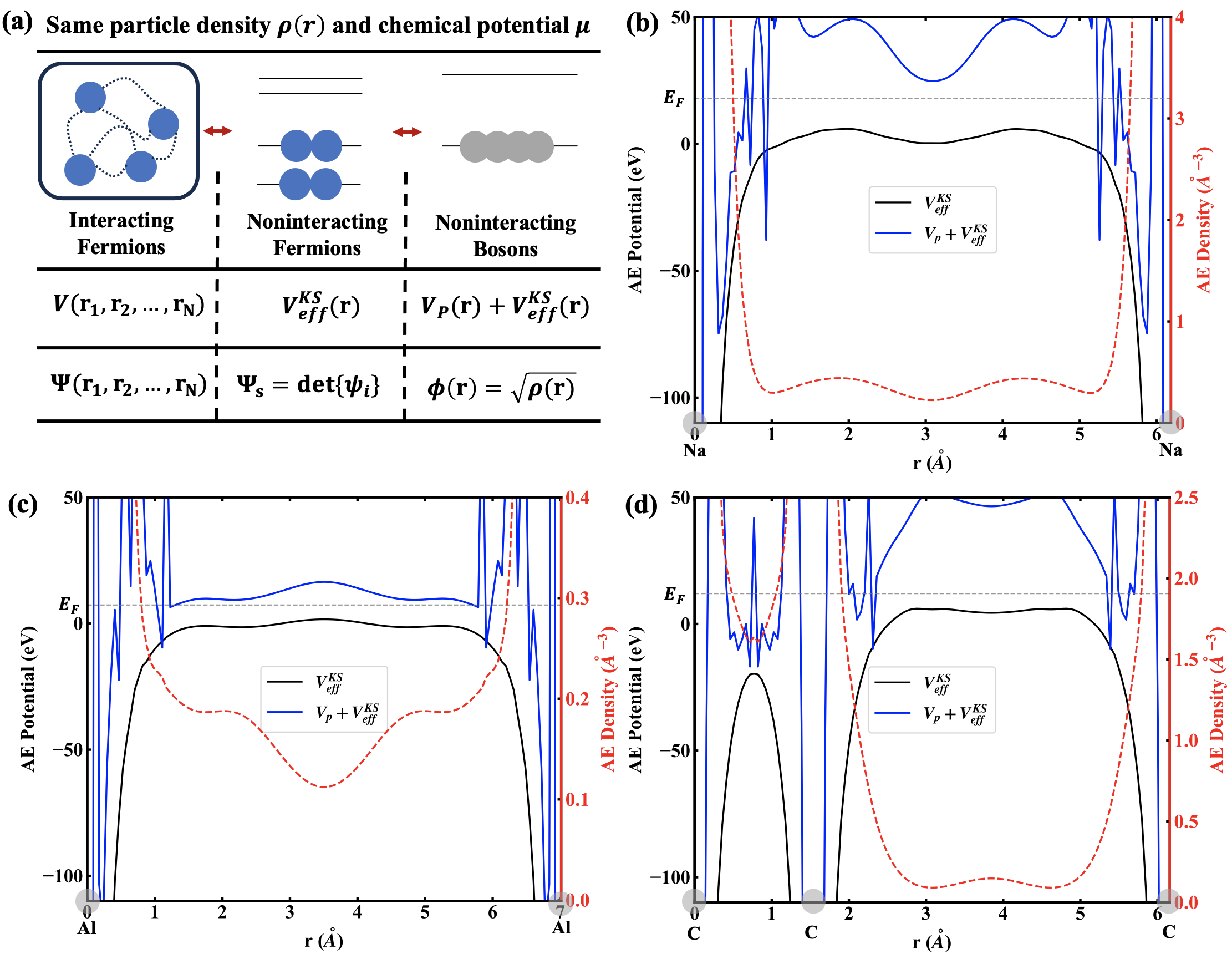}
\caption{(a) The interacting fermion systems and reference noninteracting systems. The all-electron KSDFT calculations of effective potentials and densities for (b) Na-hP4 along [1$\bar{1}1$] direction, (c) FCC-Al along [111] direction, and (d) Diamond along [111] direction.}\label{figs3}
\end{figure}

Although early theoretical models attributed the interstitial-potential well to Pauli exclusion from core electrons\cite{neaton2001constitution,rousseau2008interstitial}, no accurate or quantitative theory exists within the fermionic quasiparticle framework to describe the Pauli potential. Here, to comprehensively analyze the impact of Pauli exclusion, we mapped the noninteracting fermions (Kohn-Sham system) to the reference noninteracting bosons, where the bosons have the same charge, mass, point-like shape, chemical potential, and density as the electrons\cite{jiang2021time,xu2024recent}, as shown in Fig.~\ref{figs3}(a). In KSDFT, the electronic total energy of the interacting systems is defined as:
\beq
E[\rho]=T_{s}[\rho]+E_{H}[\rho]+E_{XC}[\rho]+E_{NE}[\rho],\label{eq:S8}
\eeq
where $E_{H}$, $E_{XC}$, and $E_{NE}$ are the Hartree energy, exchange-correlation energy, and nuclei-electron interacting energy, respectively. With the reference system of noninteracting bosons, the noninteracting kinetic energies ($T_{s}$) can be rewritten as\cite{xu2024recent}:
\beq
T_{s}[\rho]=T_{vW}[\rho]+T_P[\rho],\label{eq:S9}
\eeq
where $T_{vW}[\rho]=-\frac{1}{2}\int{\sqrt{\rho(\br)}\nabla^2\sqrt{\rho(\br)}}d^3\br$ is the von Weizs\"acker (vW) kinetic energy\cite{weizsacker1935theorie} for the noninteracting bosons occupying the single collective orbital $\phi_B(\br)=\sqrt{\rho(\br)}$ (or Madelung wavefunction\cite{madelung1927quantum}). The Pauli kinetic energy ($T_P$) is defined as the difference between the vW kinetic energy and the exact one due to the Pauli exclusion. Then, we can derive the effective potential by performing the variation on the total energy functional to be stationary with the conservation of electron number:
\begin{align}
\mu=\frac{\delta E[\rho]}{\delta\rho(\br)}=&\frac{\delta T_vW[\rho]}{\delta\rho(\br)}+\frac{\delta T_P[\rho]}{\delta\rho(\br)}+\frac{\delta (E_H[\rho]+E_{XC}[\rho]+E_{NE}[\rho])}{\delta\rho(\br)} \nonumber\\
=&V_{vW}[\rho](\br)+V_P[\rho](\br)+V_{eff}^{KS}[\rho](\br),\label{eq:S10}
\end{align}
where $V_{vW}[\rho](\br)=-\frac{1}{2}{\nabla^2\sqrt{\rho(\br)}}/{\sqrt{\rho(\br)}}$ and $V_{eff}^{KS}[\rho](\br)$ are the vW kinetic potential and Kohn-Sham effective potential. Although the exact and analytical Pauli potential ($V_P$) is still unknown, we can numerically calculate the effective potential for the reference bosons using the ground-state electron density and chemical potential calculated from KSDFT:
\beq
V_{eff}^B[\rho](\br)=V_P[\rho](\br)+V_{eff}^{KS}[\rho](\br)=\mu-V_{vW}[\rho](\br),\label{eq:S11}
\eeq
where the chemical potential is set as the Fermi energy level of KSDFT in this work ($\mu=E_F$). To include the interaction between nuclear and valence electrons, we performed the all-electron (AE) KSDFT calculations using the full-potential linearised augmented-planewave method\cite{sjostedt2000alternative} by the ELK-v10.6.2 code\cite{elk}. The AE effective potentials and electron densities are presented in Figure~\ref{figs3}(c-d) for Na-hP4 [1$\bar1$1], Al-FCC [111], and Diamond [111] directions. In this framework, the collective quantum state exhibits bound-state behavior under the all-electron effective potential containing Pauli interaction. The results completely overwhelm the behavior of fermionic quasiparticles, making it exceptionally difficult to analyze electron behavior. Therefore, the PBA effect derived from our approach offers a mathematically rigorous alternative that captures the essential physics. For the AE KSDFT calculations, we applied $18\times 18\times 11$ and $18\times 18\times 18$ $k$-grid meshes for the ground-state calculations of Na-hP4 and Al-FCC/Diamond, respectively. The potential and electron density are represented in the dense real-space grids of $120\times 120\times 120$. The eight-order finite-difference method\cite{natan2008real} is adopted for evaluating vW kinetic potential.

\bibliography{refs}

\begin{thebibliography}{10}
\urlstyle{rm}
\expandafter\ifx\csname url\endcsname\relax
  \def\url#1{\texttt{#1}}\fi
\expandafter\ifx\csname urlprefix\endcsname\relax\def\urlprefix{URL }\fi
\expandafter\ifx\csname doiprefix\endcsname\relax\def\doiprefix{DOI: }\fi
\providecommand{\bibinfo}[2]{#2}
\providecommand{\eprint}[2][]{\url{#2}}

\bibitem{sterling2024chemical}
\bibinfo{author}{Sterling, A.~J.}, \bibinfo{author}{Levine, D.~S.},
  \bibinfo{author}{Aldossary, A.} \& \bibinfo{author}{Head-Gordon, M.}
\newblock \bibinfo{journal}{\bibinfo{title}{Chemical bonding and the role of
  node-induced electron confinement}}.
\newblock {\emph{\JournalTitle{Journal of the American Chemical Society}}}
  \textbf{\bibinfo{volume}{146}}, \bibinfo{pages}{9532--9543}
  (\bibinfo{year}{2024}).

\bibitem{rousseau2008interstitial}
\bibinfo{author}{Rousseau, B.} \& \bibinfo{author}{Ashcroft, N.}
\newblock \bibinfo{journal}{\bibinfo{title}{Interstitial electronic
  localization}}.
\newblock {\emph{\JournalTitle{Physical Review Letters}}}
  \textbf{\bibinfo{volume}{101}}, \bibinfo{pages}{046407}
  (\bibinfo{year}{2008}).

\bibitem{miao2014high}
\bibinfo{author}{Miao, M.-S.} \& \bibinfo{author}{Hoffmann, R.}
\newblock \bibinfo{journal}{\bibinfo{title}{High pressure electrides: a
  predictive chemical and physical theory}}.
\newblock {\emph{\JournalTitle{Accounts of chemical research}}}
  \textbf{\bibinfo{volume}{47}}, \bibinfo{pages}{1311--1317}
  (\bibinfo{year}{2014}).

\bibitem{mao2015high}
\bibinfo{author}{Miao, M.-S.} \& \bibinfo{author}{Hoffmann, R.}
\newblock \bibinfo{journal}{\bibinfo{title}{High-pressure electrides: The
  chemical nature of interstitial quasiatoms}}.
\newblock {\emph{\JournalTitle{Journal of the American Chemical Society}}}
  \textbf{\bibinfo{volume}{137}}, \bibinfo{pages}{3631--3637}
  (\bibinfo{year}{2015}).

\bibitem{racioppi2023electride}
\bibinfo{author}{Racioppi, S.}, \bibinfo{author}{Storm, C.~V.},
  \bibinfo{author}{McMahon, M.~I.} \& \bibinfo{author}{Zurek, E.}
\newblock \bibinfo{journal}{\bibinfo{title}{On the electride nature of
  na-hp4}}.
\newblock {\emph{\JournalTitle{Angewandte Chemie International Edition}}}
  \textbf{\bibinfo{volume}{62}}, \bibinfo{pages}{e202310802}
  (\bibinfo{year}{2023}).

\bibitem{langmuir1919arrangement}
\bibinfo{author}{Langmuir, I.}
\newblock \bibinfo{journal}{\bibinfo{title}{The arrangement of electrons in
  atoms and molecules.}}
\newblock {\emph{\JournalTitle{Journal of the American Chemical Society}}}
  \textbf{\bibinfo{volume}{41}}, \bibinfo{pages}{868--934}
  (\bibinfo{year}{1919}).

\bibitem{landers1981temperature}
\bibinfo{author}{Landers, J.~S.}, \bibinfo{author}{Dye, J.~L.},
  \bibinfo{author}{Stacy, A.} \& \bibinfo{author}{Sienko, M.}
\newblock \bibinfo{journal}{\bibinfo{title}{Temperature-dependent electron spin
  interactions in lithium [2.1. 1] cryptate electride powders and films}}.
\newblock {\emph{\JournalTitle{Journal of Physical Chemistry}}}
  \textbf{\bibinfo{volume}{85}}, \bibinfo{pages}{1096--1099}
  (\bibinfo{year}{1981}).

\bibitem{lee2013dicalcium}
\bibinfo{author}{Lee, K.}, \bibinfo{author}{Kim, S.~W.}, \bibinfo{author}{Toda,
  Y.}, \bibinfo{author}{Matsuishi, S.} \& \bibinfo{author}{Hosono, H.}
\newblock \bibinfo{journal}{\bibinfo{title}{Dicalcium nitride as a
  two-dimensional electride with an anionic electron layer}}.
\newblock {\emph{\JournalTitle{Nature}}} \textbf{\bibinfo{volume}{494}},
  \bibinfo{pages}{336--340} (\bibinfo{year}{2013}).

\bibitem{dong2017electrides}
\bibinfo{author}{Dong, X.} \& \bibinfo{author}{Oganov, A.~R.}
\newblock \bibinfo{title}{Electrides and their high-pressure chemistry}.
\newblock In \emph{\bibinfo{booktitle}{Correlations in Condensed Matter under
  Extreme Conditions: A tribute to Renato Pucci on the occasion of his 70th
  birthday}}, \bibinfo{pages}{69--84} (\bibinfo{publisher}{Springer},
  \bibinfo{year}{2017}).

\bibitem{irifune2003ultrahard}
\bibinfo{author}{Irifune, T.}, \bibinfo{author}{Kurio, A.},
  \bibinfo{author}{Sakamoto, S.}, \bibinfo{author}{Inoue, T.} \&
  \bibinfo{author}{Sumiya, H.}
\newblock \bibinfo{journal}{\bibinfo{title}{Ultrahard polycrystalline diamond
  from graphite}}.
\newblock {\emph{\JournalTitle{Nature}}} \textbf{\bibinfo{volume}{421}},
  \bibinfo{pages}{599--600} (\bibinfo{year}{2003}).

\bibitem{matsuishi2003high}
\bibinfo{author}{Matsuishi, S.} \emph{et~al.}
\newblock \bibinfo{journal}{\bibinfo{title}{High-density electron anions in a
  nanoporous single crystal:[ca$_24$a$_l28$o$_64$]4+(4e-)}}.
\newblock {\emph{\JournalTitle{Science}}} \textbf{\bibinfo{volume}{301}},
  \bibinfo{pages}{626--629} (\bibinfo{year}{2003}).

\bibitem{miyakawa2007superconductivity}
\bibinfo{author}{Miyakawa, M.} \emph{et~al.}
\newblock \bibinfo{journal}{\bibinfo{title}{Superconductivity in an inorganic
  electride 12cao$\cdot$7al2o3: e$^-$}}.
\newblock {\emph{\JournalTitle{Journal of the American Chemical Society}}}
  \textbf{\bibinfo{volume}{129}}, \bibinfo{pages}{7270--7271}
  (\bibinfo{year}{2007}).

\bibitem{park2018first}
\bibinfo{author}{Park, C.}, \bibinfo{author}{Kim, S.~W.} \&
  \bibinfo{author}{Yoon, M.}
\newblock \bibinfo{journal}{\bibinfo{title}{First-principles prediction of new
  electrides with nontrivial band topology based on one-dimensional building
  blocks}}.
\newblock {\emph{\JournalTitle{Physical Review Letters}}}
  \textbf{\bibinfo{volume}{120}}, \bibinfo{pages}{026401}
  (\bibinfo{year}{2018}).

\bibitem{rioux1997kinetic}
\bibinfo{author}{Rioux, F.}
\newblock \bibinfo{journal}{\bibinfo{title}{Kinetic energy and the covalent
  bond in h2+}}.
\newblock {\emph{\JournalTitle{Chemical Educator}}}
  \textbf{\bibinfo{volume}{2}}, \bibinfo{pages}{1--14} (\bibinfo{year}{1997}).

\bibitem{nordholm2020basics}
\bibinfo{author}{Nordholm, S.} \& \bibinfo{author}{Bacskay, G.~B.}
\newblock \bibinfo{journal}{\bibinfo{title}{The basics of covalent bonding in
  terms of energy and dynamics}}.
\newblock {\emph{\JournalTitle{Molecules}}} \textbf{\bibinfo{volume}{25}},
  \bibinfo{pages}{2667} (\bibinfo{year}{2020}).

\bibitem{levine2020clarifying}
\bibinfo{author}{Levine, D.~S.} \& \bibinfo{author}{Head-Gordon, M.}
\newblock \bibinfo{journal}{\bibinfo{title}{Clarifying the quantum mechanical
  origin of the covalent chemical bond}}.
\newblock {\emph{\JournalTitle{Nature Communications}}}
  \textbf{\bibinfo{volume}{11}}, \bibinfo{pages}{4893} (\bibinfo{year}{2020}).

\bibitem{martin2022role}
\bibinfo{author}{Mart{\'\i}n~Pend{\'a}s, {\'A}.} \& \bibinfo{author}{Francisco,
  E.}
\newblock \bibinfo{journal}{\bibinfo{title}{The role of references and the
  elusive nature of the chemical bond}}.
\newblock {\emph{\JournalTitle{Nature Communications}}}
  \textbf{\bibinfo{volume}{13}}, \bibinfo{pages}{3327} (\bibinfo{year}{2022}).

\bibitem{bacskay2022orbital}
\bibinfo{author}{Bacskay, G.~B.}
\newblock \bibinfo{journal}{\bibinfo{title}{Orbital contraction and covalent
  bonding}}.
\newblock {\emph{\JournalTitle{Journal of Chemical Physics}}}
  \textbf{\bibinfo{volume}{156}}, \bibinfo{pages}{204122}
  (\bibinfo{year}{2022}).

\bibitem{ma2009transparent}
\bibinfo{author}{Ma, Y.} \emph{et~al.}
\newblock \bibinfo{journal}{\bibinfo{title}{Transparent dense sodium}}.
\newblock {\emph{\JournalTitle{Nature}}} \textbf{\bibinfo{volume}{458}},
  \bibinfo{pages}{182--185} (\bibinfo{year}{2009}).

\bibitem{kohn1965self}
\bibinfo{author}{Kohn, W.} \& \bibinfo{author}{Sham, L.~J.}
\newblock \bibinfo{journal}{\bibinfo{title}{Self-consistent equations including
  exchange and correlation effects}}.
\newblock {\emph{\JournalTitle{Physical Review}}}
  \textbf{\bibinfo{volume}{140}}, \bibinfo{pages}{A1133}
  (\bibinfo{year}{1965}).

\bibitem{neaton2001constitution}
\bibinfo{author}{Neaton, J.} \& \bibinfo{author}{Ashcroft, N.}
\newblock \bibinfo{journal}{\bibinfo{title}{On the constitution of sodium at
  higher densities}}.
\newblock {\emph{\JournalTitle{Physical Review Letters}}}
  \textbf{\bibinfo{volume}{86}}, \bibinfo{pages}{2830} (\bibinfo{year}{2001}).

\bibitem{liu2025mechanism}
\bibinfo{author}{Liu, P.}, \bibinfo{author}{Zhuang, Q.}, \bibinfo{author}{Xu,
  Q.}, \bibinfo{author}{Cui, T.} \& \bibinfo{author}{Liu, Z.}
\newblock \bibinfo{journal}{\bibinfo{title}{Mechanism of high-temperature
  superconductivity in compressed h2-molecular-type hydride}}.
\newblock {\emph{\JournalTitle{Science Advances}}}
  \textbf{\bibinfo{volume}{11}}, \bibinfo{pages}{eadt9411},
  \doiprefix\url{10.1126/sciadv.adt9411} (\bibinfo{year}{2025}).

\bibitem{pauling1931nature}
\bibinfo{author}{Pauling, L.}
\newblock \bibinfo{journal}{\bibinfo{title}{The nature of the chemical bond.
  application of results obtained from the quantum mechanics and from a theory
  of paramagnetic susceptibility to the structure of molecules}}.
\newblock {\emph{\JournalTitle{Journal of the American Chemical Society}}}
  \textbf{\bibinfo{volume}{53}}, \bibinfo{pages}{1367--1400}
  (\bibinfo{year}{1931}).

\bibitem{qiu2022superconductivity}
\bibinfo{author}{Qiu, X.-L.}, \bibinfo{author}{Zhang, J.-F.},
  \bibinfo{author}{Yang, H.-C.}, \bibinfo{author}{Lu, Z.-Y.} \&
  \bibinfo{author}{Liu, K.}
\newblock \bibinfo{journal}{\bibinfo{title}{Superconductivity in monolayer
  ba$_2$n electride: First-principles study}}.
\newblock {\emph{\JournalTitle{Physical Review B}}}
  \textbf{\bibinfo{volume}{105}}, \bibinfo{pages}{165101}
  (\bibinfo{year}{2022}).

\bibitem{wang2023crystal}
\bibinfo{author}{Wang, G.} \emph{et~al.}
\newblock \bibinfo{journal}{\bibinfo{title}{Crystal structures and
  physicochemical properties of be$_2$n and mg$_2$n as electride materials}}.
\newblock {\emph{\JournalTitle{Physical Review Applied}}}
  \textbf{\bibinfo{volume}{19}}, \bibinfo{pages}{034014}
  (\bibinfo{year}{2023}).

\bibitem{you2022emergent}
\bibinfo{author}{You, J.-Y.}, \bibinfo{author}{Gu, B.}, \bibinfo{author}{Su,
  G.} \& \bibinfo{author}{Feng, Y.~P.}
\newblock \bibinfo{journal}{\bibinfo{title}{Emergent kagome electrides}}.
\newblock {\emph{\JournalTitle{Journal of the American Chemical Society}}}
  \textbf{\bibinfo{volume}{144}}, \bibinfo{pages}{5527--5534}
  (\bibinfo{year}{2022}).

\bibitem{wang2022pressure}
\bibinfo{author}{Wang, X.} \emph{et~al.}
\newblock \bibinfo{journal}{\bibinfo{title}{Pressure stabilized
  lithium-aluminum compounds with both superconducting and superionic
  behaviors}}.
\newblock {\emph{\JournalTitle{Physical Review Letters}}}
  \textbf{\bibinfo{volume}{129}}, \bibinfo{pages}{246403}
  (\bibinfo{year}{2022}).

\bibitem{zhang2023superconductivity}
\bibinfo{author}{Zhang, X.} \emph{et~al.}
\newblock \bibinfo{journal}{\bibinfo{title}{Superconductivity in li$_8$au
  electride}}.
\newblock {\emph{\JournalTitle{Physical Review B}}}
  \textbf{\bibinfo{volume}{107}}, \bibinfo{pages}{L100501}
  (\bibinfo{year}{2023}).

\bibitem{guan2025predicted}
\bibinfo{author}{Guan, Z.}, \bibinfo{author}{Cui, T.} \& \bibinfo{author}{Li,
  D.}
\newblock \bibinfo{journal}{\bibinfo{title}{Predicted superconductivity above
  100 k in electride li$_4$rh under high pressure}}.
\newblock {\emph{\JournalTitle{Physical Review Research}}}
  \textbf{\bibinfo{volume}{7}}, \bibinfo{pages}{L012077}
  (\bibinfo{year}{2025}).

\bibitem{kronig1931quantum}
\bibinfo{author}{Kronig, R. d.~L.} \& \bibinfo{author}{Penney, W.~G.}
\newblock \bibinfo{journal}{\bibinfo{title}{Quantum mechanics of electrons in
  crystal lattices}}.
\newblock {\emph{\JournalTitle{Proceedings of the Royal Society of London.
  Series A, containing Papers of a Mathematical and Physical Character}}}
  \textbf{\bibinfo{volume}{130}}, \bibinfo{pages}{499--513}
  (\bibinfo{year}{1931}).

\bibitem{simon2013oxford}
\bibinfo{author}{Simon, S.~H.}
\newblock \emph{\bibinfo{title}{The Oxford solid state basics}}
  (\bibinfo{publisher}{OUP Oxford}, \bibinfo{year}{2013}).

\bibitem{girvin2019modern}
\bibinfo{author}{Girvin, S.~M.} \& \bibinfo{author}{Yang, K.}
\newblock \emph{\bibinfo{title}{Modern condensed matter physics}}
  (\bibinfo{publisher}{Cambridge University Press}, \bibinfo{year}{2019}).

\bibitem{tipler2007modern}
\bibinfo{author}{Tipler, P.~A.} \& \bibinfo{author}{Llewellyn, R.}
\newblock \emph{\bibinfo{title}{Modern physics}} (\bibinfo{publisher}{Macmillan
  Higher Education}, \bibinfo{year}{2007}).

\bibitem{kresse1996efficient}
\bibinfo{author}{Kresse, G.} \& \bibinfo{author}{Furthm{\"u}ller, J.}
\newblock \bibinfo{journal}{\bibinfo{title}{Efficient iterative schemes for ab
  initio total-energy calculations using a plane-wave basis set}}.
\newblock {\emph{\JournalTitle{Physical Review B}}}
  \textbf{\bibinfo{volume}{54}}, \bibinfo{pages}{11169} (\bibinfo{year}{1996}).

\bibitem{perdew1996generalized}
\bibinfo{author}{Perdew, J.~P.}, \bibinfo{author}{Burke, K.} \&
  \bibinfo{author}{Ernzerhof, M.}
\newblock \bibinfo{journal}{\bibinfo{title}{Generalized gradient approximation
  made simple}}.
\newblock {\emph{\JournalTitle{Physical Review Letters}}}
  \textbf{\bibinfo{volume}{77}}, \bibinfo{pages}{3865} (\bibinfo{year}{1996}).

\bibitem{blochl1994projector}
\bibinfo{author}{Bl{\"o}chl, P.~E.}
\newblock \bibinfo{journal}{\bibinfo{title}{Projector augmented-wave method}}.
\newblock {\emph{\JournalTitle{Physical Review B}}}
  \textbf{\bibinfo{volume}{50}}, \bibinfo{pages}{17953} (\bibinfo{year}{1994}).

\bibitem{monkhorst1976special}
\bibinfo{author}{Monkhorst, H.~J.} \& \bibinfo{author}{Pack, J.~D.}
\newblock \bibinfo{journal}{\bibinfo{title}{Special points for brillouin-zone
  integrations}}.
\newblock {\emph{\JournalTitle{Physical Review B}}}
  \textbf{\bibinfo{volume}{13}}, \bibinfo{pages}{5188} (\bibinfo{year}{1976}).

\bibitem{bloch1929quantenmechanik}
\bibinfo{author}{Bloch, F.}
\newblock \bibinfo{journal}{\bibinfo{title}{{\"U}ber die quantenmechanik der
  elektronen in kristallgittern}}.
\newblock {\emph{\JournalTitle{Zeitschrift f{\"u}r physik}}}
  \textbf{\bibinfo{volume}{52}}, \bibinfo{pages}{555--600}
  (\bibinfo{year}{1929}).

\bibitem{jiang2021time}
\bibinfo{author}{Jiang, K.} \& \bibinfo{author}{Pavanello, M.}
\newblock \bibinfo{journal}{\bibinfo{title}{Time-dependent orbital-free density
  functional theory: Background and pauli kernel approximations}}.
\newblock {\emph{\JournalTitle{Physical Review B}}}
  \textbf{\bibinfo{volume}{103}}, \bibinfo{pages}{245102}
  (\bibinfo{year}{2021}).

\bibitem{xu2024recent}
\bibinfo{author}{Xu, Q.}, \bibinfo{author}{Ma, C.}, \bibinfo{author}{Mi, W.},
  \bibinfo{author}{Wang, Y.} \& \bibinfo{author}{Ma, Y.}
\newblock \bibinfo{journal}{\bibinfo{title}{Recent advancements and challenges
  in orbital-free density functional theory}}.
\newblock {\emph{\JournalTitle{Wiley Interdisciplinary Reviews: Computational
  Molecular Science}}} \textbf{\bibinfo{volume}{14}}, \bibinfo{pages}{e1724}
  (\bibinfo{year}{2024}).

\bibitem{weizsacker1935theorie}
\bibinfo{author}{Weizs{\"a}cker, C.~v.}
\newblock \bibinfo{journal}{\bibinfo{title}{Zur theorie der kernmassen}}.
\newblock {\emph{\JournalTitle{Zeitschrift f{\"u}r Physik}}}
  \textbf{\bibinfo{volume}{96}}, \bibinfo{pages}{431--458}
  (\bibinfo{year}{1935}).

\bibitem{madelung1927quantum}
\bibinfo{author}{Madelung, E.}
\newblock \bibinfo{journal}{\bibinfo{title}{Quantum theory in hydrodynamical
  form}}.
\newblock {\emph{\JournalTitle{z. Phys}}} \textbf{\bibinfo{volume}{40}},
  \bibinfo{pages}{322} (\bibinfo{year}{1927}).

\bibitem{sjostedt2000alternative}
\bibinfo{author}{Sj{\"o}stedt, E.}, \bibinfo{author}{Nordstr{\"o}m, L.} \&
  \bibinfo{author}{Singh, D.}
\newblock \bibinfo{journal}{\bibinfo{title}{An alternative way of linearizing
  the augmented plane-wave method}}.
\newblock {\emph{\JournalTitle{Solid state communications}}}
  \textbf{\bibinfo{volume}{114}}, \bibinfo{pages}{15--20}
  (\bibinfo{year}{2000}).

\bibitem{elk}
\bibinfo{title}{{The Elk Code}}.
\newblock \bibinfo{howpublished}{\url{http://elk.sourceforge.net/}}.

\bibitem{natan2008real}
\bibinfo{author}{Natan, A.} \emph{et~al.}
\newblock \bibinfo{journal}{\bibinfo{title}{Real-space pseudopotential method
  for first principles calculations of general periodic and partially periodic
  systems}}.
\newblock {\emph{\JournalTitle{Physical Review B}}}
  \textbf{\bibinfo{volume}{78}}, \bibinfo{pages}{075109}
  (\bibinfo{year}{2008}).

\end{thebibliography}

\section*{Acknowledgements}

This research was supported by the National Natural Science Foundation of China under Grants Nos. 12305002, 12304021; the National Key Research and Development Program of China (Grant No. 2023YFA1406200), the Program for JLU Science and Technology Innovative Research Team. Part of the calculation was performed in the high-performance computing center of Jilin University and Ningbo University.

\section*{Author contributions statement}

Q.X., Z.L., and Y.M. conceived the project. Q.X. designed the theoretical framework and implemented the Kronig-Penney model solver. Q.X. and Z.L. performed the computer simulations. Q.X., Z.L., and Y.M. analyzed the results and wrote the manuscripts. All authors contributed to the discussion and revision of the manuscript.

\section*{Additional information}
%\subsection*{Accession codes}

\textbf{Competing interests} The authors declare no competing interests.

%\textbf{Correspondence and requests for materials} should be addressd to Qiang Xu, Zhao Liu, or Yanming Ma. 

\end{document}